\documentclass[preprint,12pt]{elsarticle}

\usepackage{amssymb}
\usepackage{amsmath}
\usepackage{graphicx}
\usepackage{xspace}
\usepackage{xcolor}
\usepackage{listings}
\usepackage{bm}
\usepackage{soul}

\biboptions{numbers,sort&compress}

\newcounter{bla}

\makeatletter
\newlength\cxx@c@height
\newlength\cxx@plus@height
\newcommand\cxx@plus@fontsize{\fontsize{1.6ex}{1.92ex}\selectfont}
\DeclareRobustCommand{\CC}{%
\settoheight\cxx@c@height{C}%
\settoheight\cxx@plus@height{\cxx@plus@fontsize+}%
\advance\cxx@c@height by -\cxx@plus@height%
\multiply\cxx@c@height by 10% div by 2.3
\divide\cxx@c@height by 23%
C\kern-.08em\raise\cxx@c@height\hbox{\cxx@plus@fontsize+\kern-.08em+}%
\xspace}
\makeatother

\definecolor{MyGreen}{rgb}{0,0.5,0}
\def\splitfirstchar#1#2\sentinel{\textbf{#1}#2}

\journal{Computer Physics Communications}

\begin{document}

\begin{frontmatter}

\title{BioEM: GPU-accelerated computing of Bayesian inference of electron microscopy images}

\author[MPI]{Pilar Cossio$^\ddagger$\fnref{prim_author}}
\author[FIAS]{David Rohr$^\ddagger$\fnref{prim_author}}
\author[RZG,LRZ]{Fabio Baruffa}
\author[RZG]{Markus Rampp}
\author[FIAS]{Volker Lindenstruth}
\author[MPI]{Gerhard Hummer\corref{author}}

\cortext[author]{Corresponding author -- \textit{E-mail address:} Gerhard.Hummer@biophys.mpg.de}
\fntext[prim_author]{Both authors contributed equally to the work.}
\address[MPI]{Department of Theoretical Biophysics, Max Planck Institute of Biophysics, 60438 Frankfurt am Main, Germany}
\address[FIAS]{Frankfurt Institute for Advanced Studies, Goethe University Frankfurt, Ruth-Moufang-Str.~1, 60438 Frankfurt, Germany}
\address[RZG]{Max Planck Computing and Data Facility, Giessenbachstr.~2, 85748 Garching, Germany}
\address[LRZ]{Leibniz Supercomputing Centre of the Bavarian Academy of Sciences and Humanities, 85748 Garching, Germany}
\begin{abstract}
In cryo-electron microscopy (EM), molecular structures are
determined from large numbers of projection images of individual
particles. To harness the full power of this single-molecule
information, we use the Bayesian inference of EM (BioEM) formalism.
By ranking structural models using posterior probabilities
calculated for individual images, BioEM in principle
addresses the challenge of working with highly dynamic or
heterogeneous systems not easily handled in traditional EM
reconstruction. However, the calculation of these posteriors for
large numbers of particles and models is computationally
demanding. Here we present highly parallelized, GPU-accelerated
computer software that performs this task efficiently. Our flexible
formulation employs CUDA, OpenMP, and MPI parallelization combined
with both CPU and GPU computing. The resulting BioEM software scales
nearly ideally both on pure CPU and on CPU+GPU architectures, thus
enabling Bayesian analysis of tens of thousands of images in a reasonable
time.
The general mathematical framework and robust algorithms are not limited
to cryo-electron microscopy but can be generalized for
electron tomography and other imaging experiments.
\end{abstract}


\end{frontmatter}

{\bf PROGRAM SUMMARY}

\begin{small}
\noindent
{\em Program Title:}
BioEM. \\
{\em Journal Reference:} \\
%Leave blank, supplied by Elsevier.
{\em Catalogue identifier:} \\
%Leave blank, supplied by Elsevier.
{\em Licensing provisions:}
GNU GPL v3. \\
{\em Programming language:}
\CC{}, CUDA. \\
{\em Operating system:}
Linux. \\
{\em RAM:}
Problem dependent, 100\,MB -- 4\,GB. \\
{\em Supplementary material:}
see online supplementary material.\\
{\em External routines/libraries:}
Boost 1.55, FFTW 3.3.3, MPI. \\
{\em Subprograms used:}
Situs 2.7.2\cite{wriggers2012conventions} routine for reading {\it .mrc} images. \\
{\em Running time:}
Problem dependent, $\sim 0.1$\,ms per parameter set. \\
{\em Distribution format:}
GIT repository or {\ttfamily zip} archive. \\
{\em Nature of problem:}
Analysis of electron microscopy images. \\
{\em Solution method:}
GPU-accelerated Bayesian inference with numerical grid sampling. \\
\end{small}

\section{Introduction}
\label{intro}
Cryo-electron microscopy has revolutionized structural biology~\cite{ResolRev_Kuelbrandt,Bai2015} providing
structures of chemical-motors, such as ATP synthase, ion-channels, and transporters, at atomic-level 
resolution~\cite{AllegrettiNAT2015,Wuaad2395,Ionchannel_Cheng,RyR1_Yan,Ribo_scheres,Gamma_sec_Scheres,Yuan:Nat:2016}.
Due to its near-native conditions and single-molecule character, cryo-EM is a powerful technique with great potential.
In fact, the structures of many biomolecules that are difficult to characterize using X-ray crystallography or nuclear magnetic resonance
have now been resolved with cryo-EM.

These advances are feasible because of new technologies with time-resolved direct electron detection
cameras~\cite{DEDcamera},  the development of novel image processing methods~\cite{xmipp3,eman2,Scheres_JMB_12,Bayesian-Scheres},
and the use of accelerated computing capacities of multi-core 
processors and hardware accelerators such as GPUs (Graphics Processing Units)~\cite{Frealign-GPU,Ritchie_GPU,Tagare_GPU,Stone_GPU}.
Direct electron detection cameras record an image 
rapidly as an ensemble of time-dependent frames (or movies), with an unprecedentedly low electron dose,
that capture the evolution of the individual particles in time~\cite{E_doseCheng}.
This time-resolution makes it possible to characterize the effects of beam-induced motion and radiation 
damage~\cite{BeamMotion_scheres,BeamCor_Cheng,AllegrettiELife}.
Novel image processing algorithms filter the contributions of each individual frame according
to a precalculated energy filter~\cite{Chiu_EMprot} or an in-situ B-factor assignment~\cite{DEDcam_Methods_Scheres},
and discard (or low-weight) the damaged frames.
However, not only single time-frames are discarded, but also many complete particle images.
Successful 3D classification algorithms group the images into classes that generate different 3D maps~\cite{Scheres_JMB_12,Bayesian-Scheres,Sigworth:2016},
and the images that determine the map with the highest resolution 
correspond, in most cases, to just a small fraction of the total.

Despite many improvements, there are still several scenarios in which cryo-EM algorithms face challenges to acquire high-resolution information.
Most methods rely on the hypothesis that all molecular orientations are sampled equally~\cite{particle_orient}.
This is not the case for some hydrophobic or large systems (of size comparable to the ice thickness), which acquire
preferred orientations due to the carbon grid~\cite{Pref_Ort2004,IIchap_2010}.
For disordered or heterogeneous complexes (having multiple binding partners and affinities)~\cite{Hoffman_2006,Weinkauf_2013},
it is difficult to obtain  sufficiently many particles to cover all orientations for each conformation, and
computationally challenging to assign a single orientation to one of the multiple configurations~\cite{Elad_08,Elmlund_09,NEwREV-ELad}.
A major difficulty lies in generating the initial 3D models that the algorithms use as initial alignment references.
Thus, despite the enormous advances, the effort to extend the reach of cryo-EM is ongoing.

In this work, we provide a computational tool to harness the power of cryo-EM as a true single-molecule technique.
Our algorithm performs a Bayesian inference of electron microscopy (BioEM)~\cite{CossioHummerJSB_2013},
accelerated by GPUs, to calculate the posterior
probability of a set of structural models given a set of experimental images.
In contrast to standard reconstruction algorithms, we perform a forward calculation,
building a realistic image from a model and compare it to an unmodified particle image using a likelihood function.
The BioEM posterior probability calculated in this way allows us to accurately rank and discriminate sets of structural models.

The paper is organized as follows:
we first describe the mathematical framework to create a calculated image from a model, and the Bayesian technique to
obtain the posterior probability for a set of particle images.
Then, we introduce the BioEM algorithm and describe its main routines as well as the parallelization scheme.
We demonstrate the performance and scalability of the BioEM program on a high-performance computing (HPC) cluster using two benchmark sets of particle images.
Lastly, we discuss the limitations of the method and point out future perspectives.

\section{Mathematical formulation}

\label{buildimage}

The core of the BioEM method~\cite{CossioHummerJSB_2013} is the
calculation of the posterior probability of an ensemble of structural
models, $m \in M$, given a set of experimental images,
$\omega \in \Omega$. The posterior is defined in the usual manner,
as the product of a prior probability for the various parameters
(e.\,g., orientations), and a likelihood function $L$ defined as the probability
of observing the measured image given the model and parameters,
$P(\mathrm{model}|\mathrm{data})\propto
\int d{\bm{\theta}}\,p(\mathrm{model},\mathrm{parameters}\;\bm{\theta})
L(\mathrm{data}|\mathrm{model}, \mathrm{parameters}\;\bm{\theta})$,
with parameters integrated out.
The only requirement on the models is that an EM
projection image can be calculated for them. Models can thus be
represented in a variety of ways, from atomic coordinates to 3D
electron density maps to simple geometric shapes. The key idea is to
perform forward calculations of 2D EM projection-image intensities
$I^{\mathrm{cal}}$ for given models and then compare them to the
observed image intensities $I^{\mathrm{obs}}_\omega$. The calculation
of the posterior takes into account the relevant factors in the
experiment (denoted by $\boldsymbol\theta$), such as the molecule
orientation, interference effects, uncertainties in the particle
center, normalization and offset in the intensities and noise.

\begin{figure}[!ht]
\begin{centering}
\includegraphics[width=\textwidth]{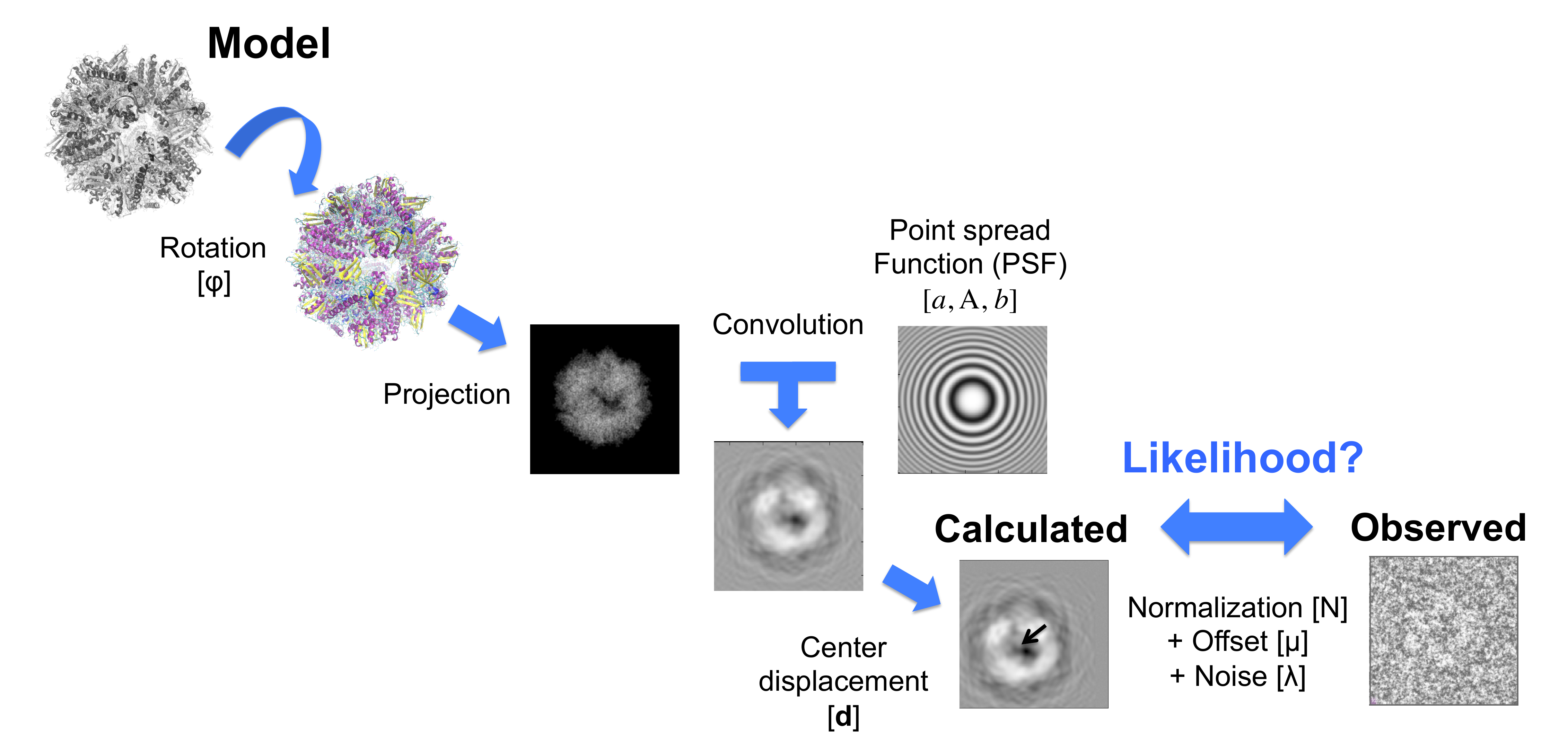}
\par\end{centering}
\caption{ {\it Steps in building a realistic image starting from a 3D model:} rotation ($\boldsymbol\varphi$),
projection, point spread function convolution ($a,A,b$), center displacement ($\mathbf{d}$), and integrated-out
parameters of normalization ($N$), offset ($\mu$) and standard deviation of the noise ($\lambda$).
The likelihood function establishes the similarity between the calculated image and the observed experimental image.}
\label{fig:likeliCons}
\end{figure}

In Fig.~\ref{fig:likeliCons}, we show a schematic representation of
how $I^{\mathrm{cal}}$ is constructed. We start from the 3D electron
density $\rho(x,y,z|m,\boldsymbol\varphi)$ of model $m$ in a
particular orientation $\bm{\varphi}$, as given by three Euler angles or
four quaternions. Under the weak-phase
approximation~\citep{int-approx}, a 2D-projection density is
determined by integration along $z$,
\begin{equation}
I^{(0)}(j,k|m,\boldsymbol\varphi)=\int_{-\infty}^\infty dz
\int_{y_k-\Delta y/2}^{y_k+\Delta y/2} dy \int_{x_j-\Delta
x/2}^{x_j+\Delta x/2} dx\,\rho(x,y,z|m,\boldsymbol\varphi)~,
\label{eq:ideal}
\end{equation}
where $(x_j,y_k)$ are the 2D positions corresponding to pixel $j$, $k$
of width $\Delta x$ and $\Delta y$. The discrete Fourier transform of
an image $I$ is defined as
\begin{equation}
\label{eq:1}
\hat{I}(\bm{s})\equiv \hat{I}(l,n)\equiv\mathcal{F}(I)=\sum_{j=1}^{N_x}\sum_{k=1}^{N_y}I(j,k)e^
{2\pi  i (lj+nk)}~.
\end{equation}
Here and in the following, we interchangeably index Fourier space with
reciprocal space vectors $\bm{s}=2\pi (l/\Delta xN_x,n/\Delta yN_y)$
or the corresponding integer indices $l$ and $n$, whichever is more convenient
for the operation at hand. Analogously, in real space we use either
physical positions $\bm{r}=(x,y)=(j\Delta x,k\Delta y)$ or the corresponding indices
$j$, $k$. $N_x$ and $N_y$ are the numbers of pixels in $x$ and $y$
directions, respectively.

The inverse Fourier transform is
\begin{equation}
\label{eq:2}
I(j,k)\equiv \mathcal{F}^{-1}(\hat{I})=\frac{1}{N_\mathrm{pix}}\sum_{l=1}^{N_x}\sum_{n=1}^{N_y}\hat{I}(l,n)e^{-2\pi
i (lj+nk)}~,
\end{equation}
where $N_\mathrm{pix}=N_xN_y$ is the total number of pixels.

We account for interference and inelastic scattering effects in EM
imaging through the contrast transfer function (CTF) and the envelop
function (Env), respectively, which are both assumed to be radially
dependent for simplicity. The product of CTF and Env is the
Fourier-space equivalent of the real-space point spread function
(PSF),
\begin{equation}
\label{eq:3}
\mathrm{CTF}(s) \mathrm{Env}(s)=\mathcal{F}(\mathrm{PSF})~,
\end{equation}
where $s=|\bm{s}|$. As functional
forms, we assume
$\mathrm{CTF}(s|a,A)=-A\cos(as^2/2)-\sqrt{1-A^2}\sin(as^2/2)$ and
$\mathrm{Env}(s|b)=e^{-bs^2/2}$ with coefficients $a$, $A$, and $b$
\cite{Penczek_2010}.

In Fourier space, the interference effects on the ideal image are then accounted by
\begin{equation}
\label{eq:4}
\hat{I}^{(1)}(\bm{s}|m,\boldsymbol\varphi,a,A,b)=\hat{I}^{(0)}(\bm{s}|m,\boldsymbol\varphi) \mathrm{CTF}(s|a,A)
\mathrm{Env}(s|b)~,
\end{equation}
which corresponds to a convolution with the PSF in real space,
\begin{equation}
\label{eq:5}
I^{(1)}(\bm{r}|m,\boldsymbol\varphi,a,A,b)= \sum_{\bm{r}'}
I^{(0)}(\bm{r}'|m,\boldsymbol\varphi)\mathrm{PSF}(|\bm{r}-\bm{r}'||a,A,b)~.
\end{equation}
Implicit in this procedure is the assumption that the pixel size is
small compared to significant variations in the PSF.

The calculated image also accounts for uncertainties in the particle position, and variations in the imaging conditions
\begin{equation}
I^{\mathrm{cal}}(\bm{r}|m,\boldsymbol\varphi,a,A,b,\mathbf{d},N,\mu) =N I^{(1)}(\bm{r}+\mathbf{d}|m,\boldsymbol\varphi,a,A,b)-\mu~,
\end{equation}
where $\mathbf{d}=(d_x,d_y)$ is a translation vector shifting the image
and thus the particle center pixel by pixel, $N$ scales the intensity,
and $\mu$ is an intensity offset. The likelihood function
$L(\omega|m,\boldsymbol\theta)$ of model $m$ with image parameters
$\boldsymbol\theta=(\boldsymbol\varphi,a,A,b,\mathbf{d},N,\mu,\lambda)$,
establishes a measure of similarity between $I^{\mathrm{cal}}$
and $I^{\mathrm{obs}}_\omega$,
\begin{equation}
L(\omega|m,\boldsymbol\theta)=\frac{\exp\left(-{{\sum_{j,k}\left[I_{\omega}^{\mathrm{obs}}(j,k)-I^{\mathrm{cal}}(j,k|m,\boldsymbol\varphi,a,A,b,\mathbf{d},N,\mu)\right]^{2}}}/{2\lambda^{2}}\right)} {
\left(2\pi\lambda^2\right)^{N_\mathrm{pix}/2}}
~.\label{eq:likeli}
\end{equation}
Here we
assumed for simplicity uncorrelated Gaussian noise in each pixel with
zero mean and standard deviation $\lambda$.

The likelihood function, $L(\omega | m, \boldsymbol\theta)$, measures the agreement between individual
observed and calculated images for fixed parameters $\boldsymbol\theta$.
In contrast to  maximum likelihood techniques~\cite{Rev_ML-10} that determine a single optimal parameter set,
we perform a Bayesian analysis covering a wide range of possible parameter sets.
This range, and the weight of individual sets, is defined by the prior probability $p(\boldsymbol\theta)$ 
(see the Supplementary Information for details about the priors implemented in the code), which is
combined with the prior of the model, $p_{M}(m)$. The
Bayesian posterior probability of a model, $m$, given an image, $\omega$, is then a weighted integral over the product of prior and likelihood,
\begin{equation}
P_{m\omega} \propto \int
L(\omega|m,\boldsymbol\theta)p_M(m)p(\boldsymbol\theta)
d\boldsymbol\theta~.
\label{eq:Pmom}
\end{equation}

The posterior probability of an ensemble containing multiple models ($m \in M$) with relative weights $w_m$ (where $\sum_mw_m=1$), given a set of images ($\omega \in \Omega$), is then
\begin{equation}
P(M|\Omega)  \propto \prod_{\omega=1}^{\Omega}\sum_{m=1}^{M} w_m
P_{m\omega}~,
\label{eq:pb2}
\end{equation}
i.\,e., the product over independent images of weighted sum over ensembles.

\section{Algorithm and Optimization}

\label{sec:optimizations}

To evaluate the posterior probabilities $P_{m\omega}$ in
Eq.~\ref{eq:pb2} for every model $m \in M$ and image
$\omega \in \Omega$, we have to compute the integral in
Eq.~\ref{eq:Pmom} over the parameters
$\boldsymbol\varphi,a,A,b,\mathbf{d},N,\mu$, and
$\lambda$. As shown in ref.~\cite{CossioHummerJSB_2013}, we can
integrate noise $\lambda$, normalization $N$, and offset $\mu$
analytically. The resulting analytical expression of the posterior
probability
$P_{m\omega}(\boldsymbol \varphi,a,A,b,\mathbf{d})$ as a
function of the remaining parameters is shown in Eq.~S8 of the Supplementary
Information. Average and mean squared averages of the intensities
of the observed and
calculated images (Eqs.~S3 - S6) can be precomputed.
Additionally, Eq.~S8 involves the estimation of the cross-correlation
between the calculated and observed images, $C_{co}$ from Eq.~S7. The
remaining integrals in Eq.~\ref{eq:Pmom} over orientations parametrized by
$\boldsymbol\varphi$, the PSF with parameters ($a,A,b$), and center displacements $\mathbf{d}$ are evaluated
numerically.
Importantly, for each parameter combination (with the possible exception of
$\mathbf{d}$; see below), the cross-correlation between the
$I^{\mathrm{cal}}$ and $I^\mathrm{obs}$ has to be calculated in a
computationally demanding step.

The data dependency in the construction of the calculated image is sketched in Fig.~\ref{fig:likeliCons}. The model must
be first rotated, then projected, then convoluted, and then displaced.
The loop over the experimental images is, in principle, independent but it is best to nest this loop inside the loop over the model rotations and PSF convolutions.
Then, the same rotation, projection, and convolution do not have to be
computed repeatedly for every observed image.

Listing~\ref{list:code1} presents a pseudo-code of the BioEM algorithm.
Here, the subroutine {\ttfamily\footnotesize compute\_probability} computes the posterior probability from Eq.~S8, and {\ttfamily\footnotesize norm\_images[]} is an array with precomputed~$C_o$ and $C_{oo}$ (from Eqs.~S3 and S5) for all observed images.
\\
\begin{lstlisting}[frame=single,basicstyle={\ttfamily\footnotesize},label={list:code1},caption={Pseudocode of the numerical integration.}\\ ]
for (r = 0;r < num_rot;r++) {
  m_rot = do_rotation(r, model);
  m_proj = do_projection(m_rot);
  for (c = 0;c < num_convolutions;c++) {
    m_conv = do_psf_convolution(m_proj, c);
    for (d = 0;d < num_displacements;d++) {
      m_disp = do_2d_displacement(m_conv, d);
      norm_model = compute_norm(m_disp);
      for (w = 0;w < num_images;w++) {
        cross_corr = compute_cross_corr(m_disp, image[w]);
        image_probability[w] += compute_probability
          (norm_images[w], norm_model, cross_corr);
}}}}
\end{lstlisting}

In the following, we 
describe the optimizations applied during the development, including some technical details.
The measurements presented in this section have been taken on our development
system, which hosts a typical, workstation-class Intel Core i7-980 3.33\,GHz 6-core CPU and an
NVIDIA GTX Titan GPU.
This is complemented in Section~\ref{sec:perf} by a systematic study of the parallel efficiency and the
GPU acceleration, performed on a HPC cluster with
server-class CPUs and GPUs.

\subsection{Model Rotation}

The BioEM algorithm offers two representations of model orientations
in 3D space: with Euler angles or with quaternions.
Whereas the Euler angles are most commonly used in EM image processing softwares, we have found it
more suitable to use quaternions. The advantage is that with the
quaternions uniform sampling of the group {\it SO(3)}
of rotations in 3D space is more readily implemented \cite{Yershova2010}.

\subsection{Imaging Effects}

After the rotation and projection, we construct the Fourier transform
of the ideal image, $\mathcal{F}(I^{(0)})$, from Eq.~\ref{eq:ideal}.
Interference and imaging effects are taken into account by
multiplication with CTF and Env in Fourier space according to
Eq.~\ref{eq:4}, or by convolution with the PSF in real space (which
can be advantageous, e.\,g., if the PSF is zero everywhere except near
the origin). In the code, the PSFs to be sampled over, and their
Fourier transforms, are precalculated in the form of radially
symmetric 2D images, and stored in memory.

\subsection{Center Displacement}
\label{centdis}

\begin{figure}[!ht]
\begin{centering}
\includegraphics[width=0.75\textwidth]{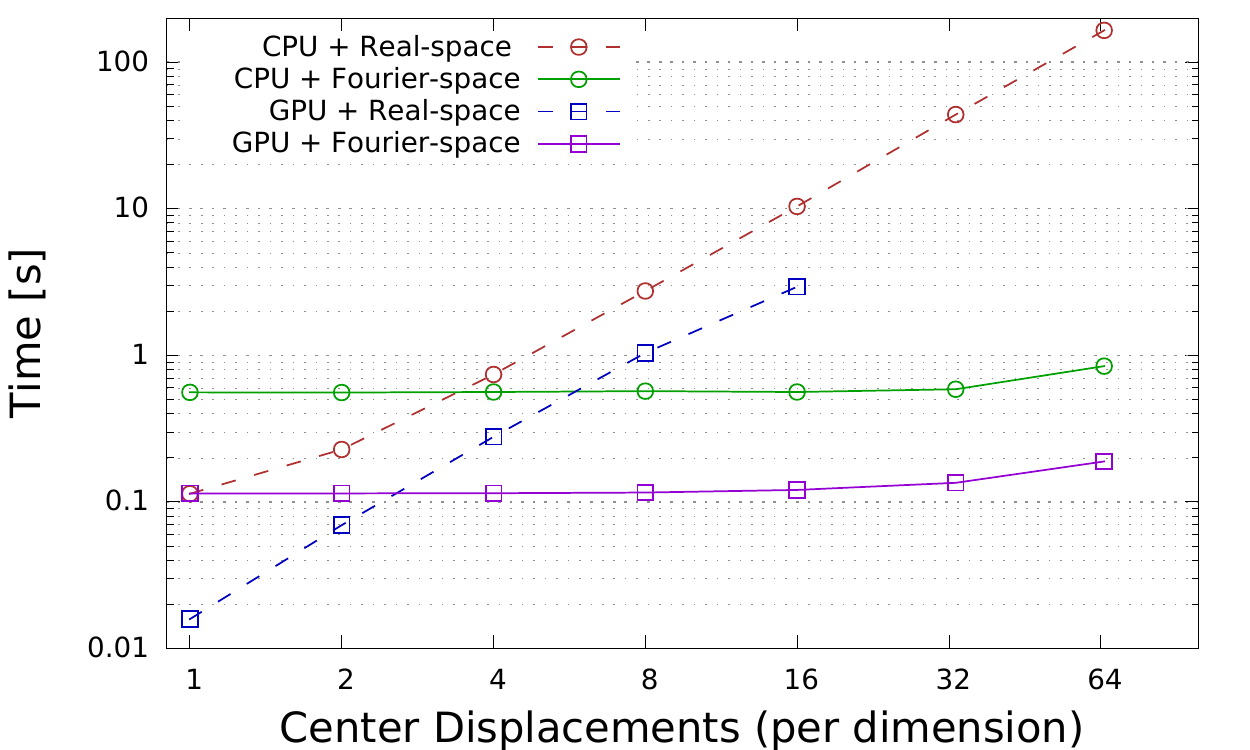}
\par\end{centering}
\caption{Processing time of the real-space (dashed) and Fourier-space (solid) variants to
compute the cross-correlation of the calculated image and the observed image, $C_{co}(\mathbf{d})$,
as a function of the number of center displacements $\mathbf{d}$ for both the CPU (circles) and GPU (squares).
The measurement was taken for the 11000-image set on the same benchmark system as specified in Table~\ref{tab:timeline}.}
\label{fig:shifts}
\end{figure}

To evaluate the integral over the image translation vectors $\mathbf{d}$
we have implemented two alternative approaches suitable for small and large sets of $\mathbf{d}$, respectively.
In a real space formulation, the cross-correlation is simply calculated for different
displacements $\mathbf{d}$. In Fourier space, we use the fact that
the cross-correlation of the calculated and the observed images
shifted by~$\mathbf{d}$ (denoted by $C_{co}(\mathbf{d})$), is given by
\begin{equation}
C_{co}(\mathbf{d})= \mathcal{F}^{-1}[\mathcal{F}(I^\mathrm{obs}) \cdot \mathcal{F}(I^{(1)})] (\mathbf{d}),
\label{eq:fourier}
\end{equation}
where $\mathcal{F}^{-1}$ is the inverse Fourier transformation and the
dot indicates the complex product of Fourier components at equal
$\bm{s}$ vectors, i.\,e.,
\begin{equation}
C_{co}(d_x,d_y)= \frac{1}{N_\mathrm{pix}}\sum_{l=1}^{N_x}\sum_{n=1}^{N_y}\hat{I}^\mathrm{obs}(l,n)\overline{\hat{I}^{(1)}(l,n)}
e^{-2\pi i (ld_x+nd_y)},
\label{eq:fourier_explicit}
\end{equation}
where the overline indicates the complex conjugate.
In this way, we can take advantage of the fact that the convolution with the PSF carried out in Fourier space produces $\hat{I}^{(1)}$, and that $\hat{I}^\mathrm{obs}$ can be precalculated.
Hence, the only computationally intense part is the Fourier backtransformation~$\mathcal{F}^{-1}$ in Eq.~\ref{eq:fourier}.
If we compare the two algorithm variants, the real-space one requires the computation of as many cross-correlations as there are displacements.
The Fourier variant requires one Fourier backtransformation, irrespective of the number of displacements.
Naturally, the first version should have advantage with few displacements, while variant two will be superior for many displacements.
In fact, Fig.~\ref{fig:shifts} shows that the Fourier version is superior in most cases, both on the CPU and the GPU.

The Fourier variant differs from the real-space version in that it wraps around the borders of the experimental image instead of just cropping it.
However, comparing the results for certain larger datasets, we can conclude that this does not pose a problem,
because the particle is generally located in the center of the image while the region around it (the borders) consists mostly of noise.
Moreover, using the Fourier variant, we have the advantage that we do not have to convert the calculated image back to real space for computing the cross-calculation, which saves some of the Fourier transforms.
We still need the constants $C_o$ and $C_{oo}$, but these can be easily computed in Fourier space using  Plancherel's theorem,
e.\,g., $C_{oo} \propto \sum |\mathcal{F}(I^\mathrm{obs})|^2$.

We perform the Fourier-transformation using the Fast Fourier Transform (FFT) libraries~\cite{FFTW05} {\ttfamily fftw} on the CPU and {\ttfamily cuFFT} on the GPU.
Since all images (calculated and observed) are real, we use the fast real-Hermitian FFT variant offered by these libraries.
The computing time of the FFT algorithms depends on the image size (see Supplementary Fig.~S1). Box-sizes that are powers of 2 per dimension
or follow standard EM/FFT suggestions~\cite{Tang:JSB:2007} have the best performance.

\subsection{Numerical Precision}
We computed the numerical integral in Eq.~\ref{eq:Pmom} using single and double precision with and without the
the Kahan summation algorithm~\cite{kahan:1965}. We found that all options lead to the same numerical result, mainly, because
the likelihood function is sharply peaked around its maxima and few parameter sets
contribute to the summation. Thus, we set as default the fastest setup which uses single precision without the Kahan summation.
All benchmarks shown in this paper are based on this single-precision setting.

\subsection{Blocking}

A common technique to optimize memory access patterns of nested loops (for instance for matrix operations) is blocking, which improves the cache usage.
Alternatively, if the memory access pattern does not (or almost not) depend on one of the loops, one can place that loop as the innermost loop.
We apply this optimization for the real-space variant.
The loop over the displacements shifts the memory access only by one entry per iteration, so we use this loop as the
innermost loop (i.\,e., exchange the loops over~{\ttfamily w} and~{\ttfamily d} in Listing~\ref{list:code1}).
Since, for the Fourier variant, the FFTs are computationally dominant, we do not need the blocking optimization in this case.

\subsection{Vectorization}
\label{vect}

For the real-space version, we have written the code to support compiler auto vectorization.
By checking the disassembly of the object file we verified that the compiler vectorized the code exactly in the way we intend.
Since optimized FFT libraries use vectorization, we do not have to take action in this respect for the Fourier variant.

\subsection{Parallelization}

In order to speed up processing of the very compute-intense BioEM task, we parallelize the processing on top of the vectorization.
We consider parallelization over the cores inside one compute node, parallelization over multiple compute nodes, and usage of parallel accelerator devices like GPUs to speed up processing.

\begin{figure}[!ht]
\begin{centering}
\includegraphics[width=\textwidth]{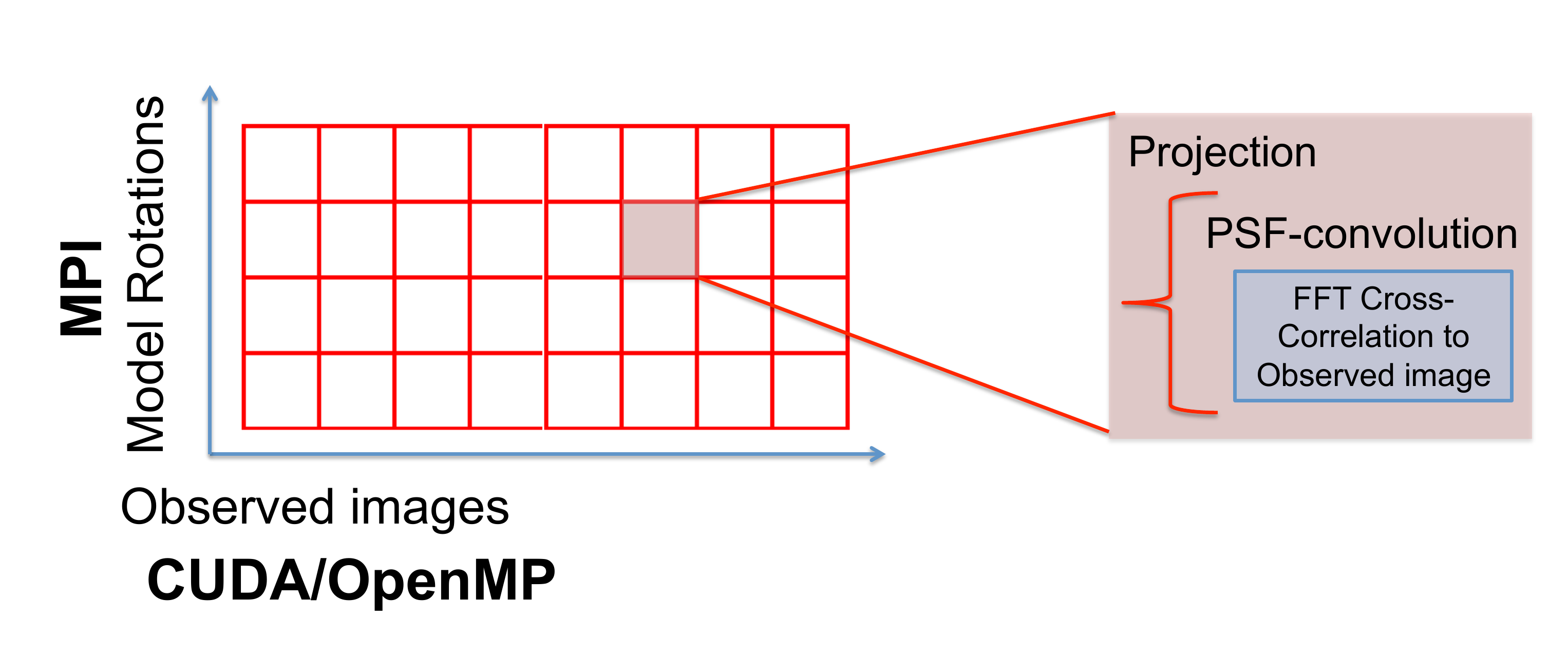}
\par\end{centering}
\caption{Representation of the double parallelization scheme used in the BioEM algorithm.
MPI is used for parallelization over different model rotations while CUDA and/or OpenMP is used for parallelization over different images.
For each pair of model orientation and particle image (zoomed box), a loop over the PSF convolution kernels is performed, 
and the cross-correlation to the observed image is calculated using a fast-Fourier algorithm.}
\label{fig:BioEM_algo}
\end{figure}

\subsubsection{CPU Usage}

It is desirable to parallelize the inner loop over the images on a shared memory architecture.
This way, we need to compute rotation, projection, and convolution only once, and we can then 
reuse the calculated image~$I^{\mathrm{cal}}$ for comparison with all observed images~$I^{\mathrm{obs}}$.
We employ OpenMP to process the comparison of~$I^{\mathrm{cal}}$ to all images in parallel.
The outermost loop over the rotations does not have any dependencies (except for the source data), 
hence we have chosen to parallelize it via the Message Passing Interface (MPI) to support utilization of many compute nodes in parallel.
Fig.~\ref{fig:BioEM_algo} shows a schematic representation of the parallelization approach.
Each grid cell represents a single rotation to a single experimental image, inside which the 
integrals over the projection, PSF convolution, and center displacement are performed (zoomed square Fig.~\ref{fig:BioEM_algo}).
The Fourier algorithm calculates the center displacement and cross-correlation to the experimental image simultaneously.
If there are only a few experimental images, then also the rotation, projection, and Fourier transformation of $I^\mathrm{cal}$ 
take a non negligible time (up to~$30$\,\% for the small dataset from the Results section).
In order to speed this up, BioEM can use OpenMP to precalculate all Fourier transformed images~$I^{\mathrm{cal}}$ in parallel.

\begin{figure}[!ht]
\begin{centering}
\includegraphics[width=0.75\textwidth]{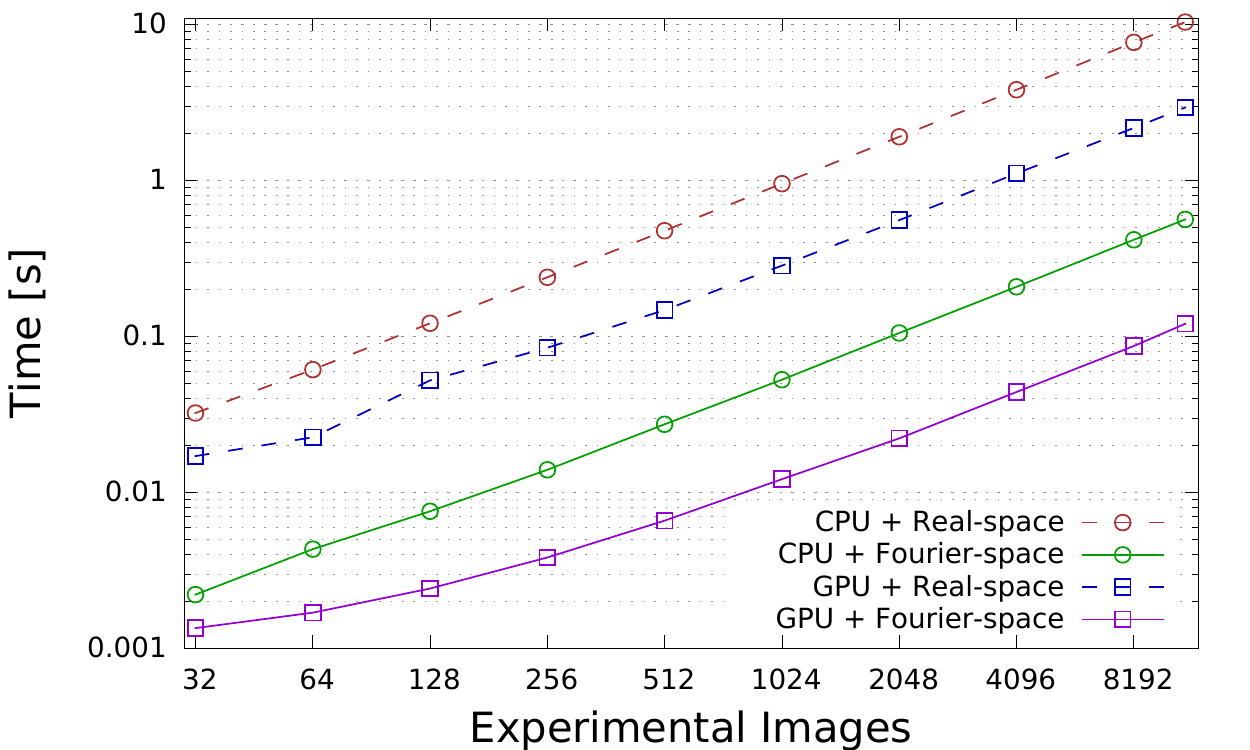}
\par\end{centering}
\caption{Processing time of the real-space and the Fourier variants for the calculation of the posterior probability, as a function of the number of observed images.
Results are shown for the CPU (circles), and the GPU (squares) for both variants real-space (dashed) and Fourier-space (solid).
The measurement was taken for the 11000-image set on the same benchmark system as specified in Table~\ref{tab:timeline}.}
\label{fig:maps}
\end{figure}

\subsubsection{GPU Usage}

In recent years, Graphics Processing Units (GPUs) have shown significant speedup in many applications, among them the FFT, which is heavily used by BioEM.
In order to leverage this potential, we have adapted BioEM to run on GPUs.
BioEM can use CUDA for the cross-correlation step, which essentially consists of an image multiplication in Fourier space and a Fourier back-transformation.
We did not consider bringing more of the steps to the GPU because the other parts are not time critical and can be processed well by the CPU.
We use a pipeline where the CPU can prepare the next rotations and convolutions while the GPU runs the comparison of the last~$I^{\mathrm{cal}}$ to all observed images.
We also arrange the remote direct memory access (RDMA) transfer of the new data from the host to the GPU in the pipeline asynchronously via CUDA streams.
In this way, BioEM keeps executing GPU kernels~100\,\% of the time and there is no GPU idle time.
In order to use both CPU and GPU to the full extent, BioEM splits the work between GPU and CPU and uses both processor types jointly for the comparison step.
Fig.~\ref{fig:maps} shows the computational time as a function of the number of the observed images. 
Except for very small sets of images, the processing time depends linearly on the number of observed images. 
Here the transition to linear scaling occurs between~128 and~256 images and is, thus, below what one would encounter in typical applications.

Table~\ref{tab:timeline} summarizes the evolution of the code's performance over
subsequent development phases, starting out from a first, serial prototype
version executed on a single CPU core (Intel Core i7, 3\,GHz) to a full, GPU-accelerated
multicore node. We would like to emphasize that the huge overall speedup (3 or 4 orders of
magnitude for different setups with 130 or 11000 images, respectively) is to a large
degree due to algorithmic optimizations pointed out in Sections~\ref{centdis} to \ref{vect}
and OpenMP parallelization of the CPU version. The latter defines the performance baseline
for a fair comparison of the GPU version which delivers a speedup by a factor
of 5 in this case (6-core CPU vs. Titan GPU).

\begin{table}[b!]
\begin{center}
\caption{Performance evolution of BioEM through subsequent optimization phases
(from top to bottom, first column)
as measured on a workstation with an Intel Nehalem Core i7-980 6-core CPU
(3.33\,GHz) and an NVIDIA GTX Titan GPU.
To test the flexibility of the code, we used two different setups (second column):
11000 images ($224 \times 224$ pixels)
of the F$_{420}$-reducing hydrogenase system~\cite{AllegrettiELife,Mills_2013}, 
and 130 images ($170 \times 170$ pixels) 
of the chaperonin GroEL~\cite{ref_3C9V}, both with one PSF, one orientation and one center displacement.
The third column states the absolute runtime and the forth and
fifth column give the relative speedup compared to the previous, or initial
code version, respectively.}
\label{tab:timeline}
\scalebox{0.825}{
\begin{tabular}{lrrrr}
\hline
Version & Number & Time & Speedup & Speedup   \\
&  of Images &      & (incremental)  & (cumulative)   \\
\hline
First prototype version     &   130  & $\approx 2$ s    & -    & -     \\
		    & 11000  & $\approx 1200$ s & -    & -     \\
\hline
First BioEM \CC      &   130  & 0.533 s          & 4.00  &     4 \\
real-space version		    & 11000  & 312.4 s          & 4.00  &     4 \\
\hline
Vectorized \CC real-space version,     &   130  & 0.083 s          & 6.39 &    26 \\
with optimizations from Section~\ref{sec:optimizations}
		    & 11000  & 42.62 s          & 7.33 &    29 \\
\hline
Fourier-space version          &   130  & 0.120 s          & 0.69 &    17 \\
		    & 11000  & 8.256 s          & 5.18 &   151 \\
\hline
Real-Hermitian FFT version  &   130  & 0.048 s          & 2.49 &    44 \\
		    & 11000  & 2.809 s          & 2.94 &   444 \\
\hline
Parallelization with        &   130  & 0.011 s          & 4.46 &   537 \\
OpenMP (6 cores)            & 11000  & 0.581 s          & 4.84 &  2151  \\
\hline
GPU usage with CUDA         &   130  & 0.00213 s        & 5.07 &   998 \\
(1 NVIDIA Titan GPU)        & 11000  & 0.106 s          & 5.48 & 11790 \\
\hline
Combined CPU / GPU          &   130  & 0.00208 s        & 1.02 &  1023 \\
usage                       & 11000  & 0.091 s          & 1.16 & 13733 \\
\hline
\end{tabular}}
\end{center}
\end{table}

\subsection{Extra Features}

In addition to calculating the posterior probability of a model given a set of experimental images, the code provides several extra features:
\begin{itemize}
\item {\it Synthetic map:} print the synthetic map corresponding to a specific parameter set, $\boldsymbol\theta$.
\item {\it Maximizing parameters:}  report the grid value parameters that give a maximum of the posterior probability.
\item {\it Posterior for orientations:} obtain the posterior probability as a function of the rotational (Euler or quaternion) angles.
\end{itemize}
Details on using each feature are provided in the user manual.

\subsection{Margin for Improvements}

Using the CUDA profiler, we measured that more than 75\% of the compute time is spent in the  {\ttfamily cuFFT} library for both datasets.
The fraction of CPU time spent in {\ttfamily fftw} is even larger.
The majority of the remaining time is used for pixel-wise multiplication of the images in Fourier space, which is by definition memory-bandwidth limited, in particular on the GPU.
It is a small inefficiency in this respect that the images are stored to memory after 
the multiplication and then read again for the FFT, but this can hardly be avoided due to the use of the FFT libraries.
The FFT libraries themselves are already well optimized, thus the margin for an additional speedup is limited.

\section{Performance}
\label{sec:perf}

This section presents a performance evaluation of the BioEM software, focusing on parallel efficiency and 
GPU performance obtainable on a typical high-performance compute cluster that is employed for production runs with BioEM.
The tests were performed on the high-performance system \emph{Hydra} of the Max-Planck-Society, 
operated by the Max-Planck Computing and Data Facility in Garching, Germany.
It consists of dual-socket nodes equipped with Intel Xeon E5-2680 v2 CPUs (20 physical cores per node with 
2 hyperthreads per core) and interconnected with a high-performance network (FDR InfiniBand).
A subset of the nodes is equipped with two NVidia K20X ``Kepler'' GPUs each.
For the benchmarks on the \emph{Hydra} system, Intel compiler suite XE 2014, CUDA 5.5 and FFTW 3.3.3 were used on top of the Linux operating system SLES11.

We selected a benchmark set with 100 (denoted by ``small") and 2000 (denoted by ``large") 
experimental particle images of the 1.2 MDa F$_{420}$-reducing hydrogenase (Frh) system~\cite{AllegrettiELife,Mills_2013}.
Each particle image had $224 \times 224$ pixels with 1.32 \AA{} of pixel size.
An all-atom structure ($\sim 82000$ atoms) built from the 3D density map~\cite{AllegrettiELife} was used as the reference model.
The numerical integrals in Eq.~\ref{eq:Pmom} were performed over grids with 13500 orientations, 64 PSFs, and 400 center displacements.
This parameter setup is consistent with a case without prior knowledge of the symmetry of the system or the orientations of the particle images. 
Thus, we did not take advantage of the 12-fold symmetry of the Frh complex to reduce the orientational search. 
However, in practical applications this can be easily implemented, by searching over a restricted set of Euler angles or corresponding quaternions.

\begin{figure}[!ht]
\begin{centering}
\includegraphics[width=0.75\textwidth]{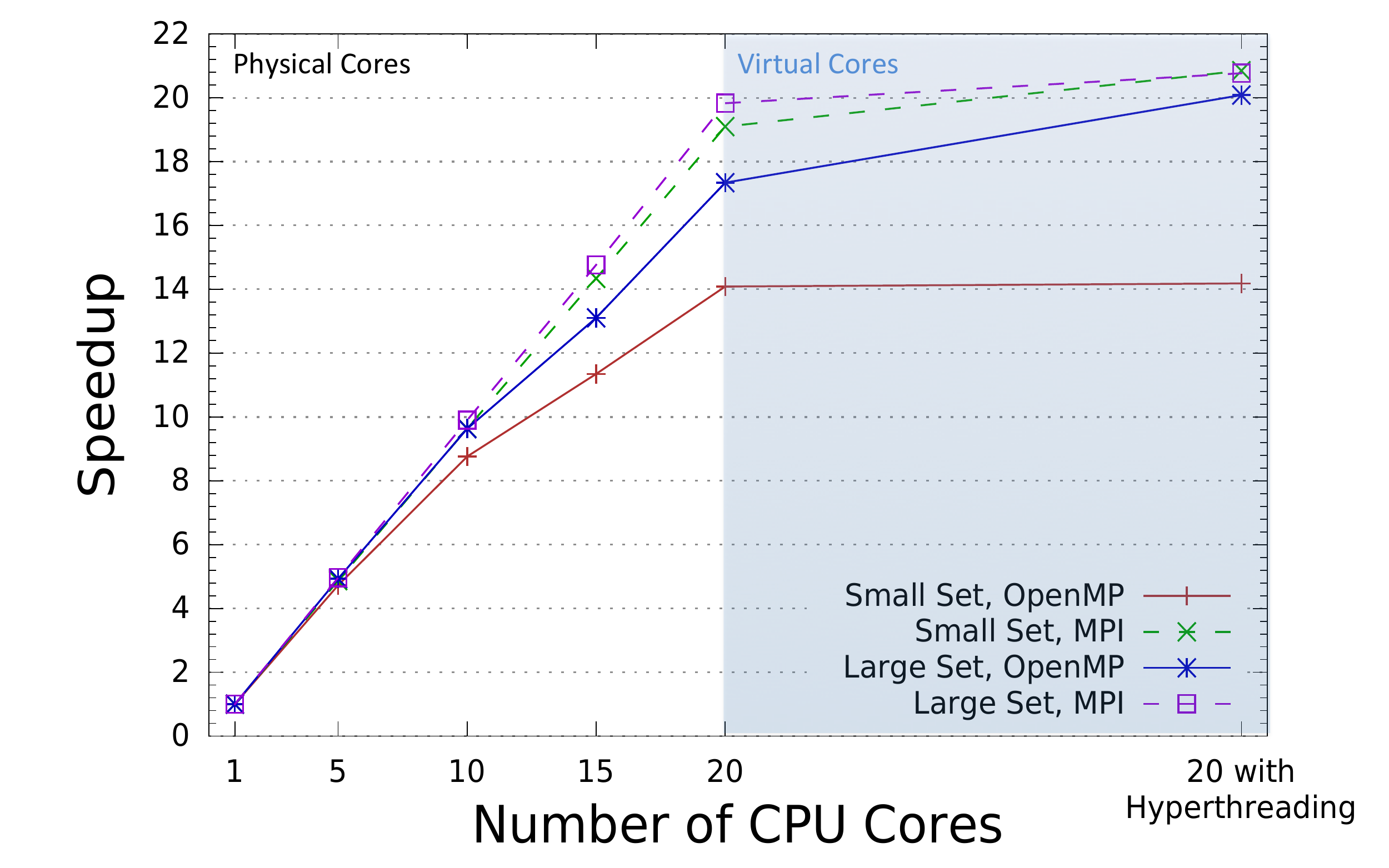}
\par\end{centering}
\caption{BioEM speedup compared to a single thread as a function of the number of employed CPU cores using the OpenMP (solid lines),
over the observed images, or MPI (dashed lines), over the orientations, parallelization for both the small and large datasets. The shaded region indicates the use of Hyper-Threading.}
\label{fig:perf_single}
\end{figure}

Fig.~\ref{fig:perf_single} shows that we can achieve almost perfect linear scaling with the number of physical  cores in a node for both datasets.
BioEM allows us to parallelize over the cores inside a node in several ways: over the observed images via OpenMP,
 or over the orientations via MPI, or with a combination of MPI and OpenMP in a hybrid setup.
The figure compares OpenMP to MPI scalability.
There are two effects that limit pure OpenMP scaling.
First, there are unavoidable non-uniform memory access (NUMA) effects because common global data
(e.\,g., the calculated images) are stored only once and thus are scattered over NUMA domains.
This becomes apparent in Fig.~\ref{fig:perf_single} which shows that both  the large (blue, solid line) and 
the small setup (red, solid line) show very good scalability up to the maximum of 10 cores of a NUMA domain.
Second, in particular for the small dataset, synchronization after the computation of the likelihood limits the performance.
In contrast, these aspects do not affect the pure MPI setup with each process mapped to an individual core.
The MPI configurations (dashed lines) exhibit nearly perfect scaling up to the maximum of 20 physical cores in the node because both 
datasets have sufficiently many orientations for MPI parallelization.
Thus, we note as a side result that memory bandwidth is not a limiting factor in this context.
Since BioEM is compute bound by the FFTs, Hyper-Threading yields a small but non-negligible improvement.

An important advantage of the OpenMP  parallelization is its smaller memory footprint.
While in the OpenMP case the threads share a copy of the observed images, each process in the MPI case has its own copy.
The large dataset, for example, requires 35~GB (1.75~GB per MPI-process), which can already be prohibitive on some of 
today's HPC clusters and the memory requirement increases further with larger datasets.
Moreover, with a plain MPI parallelization the total number of
orientations poses a strict upper limit on the number of MPI tasks.
This would ultimately limit the strong scalability of BioEM, in particular for
smaller problems with only very few orientations.
The steadily increasing number of cores per CPU and the stagnating (or
even decreasing) per-core performances will further exacerbate these constraints in the future.

\begin{figure}[!ht]
\begin{centering}
\includegraphics[width=\textwidth]{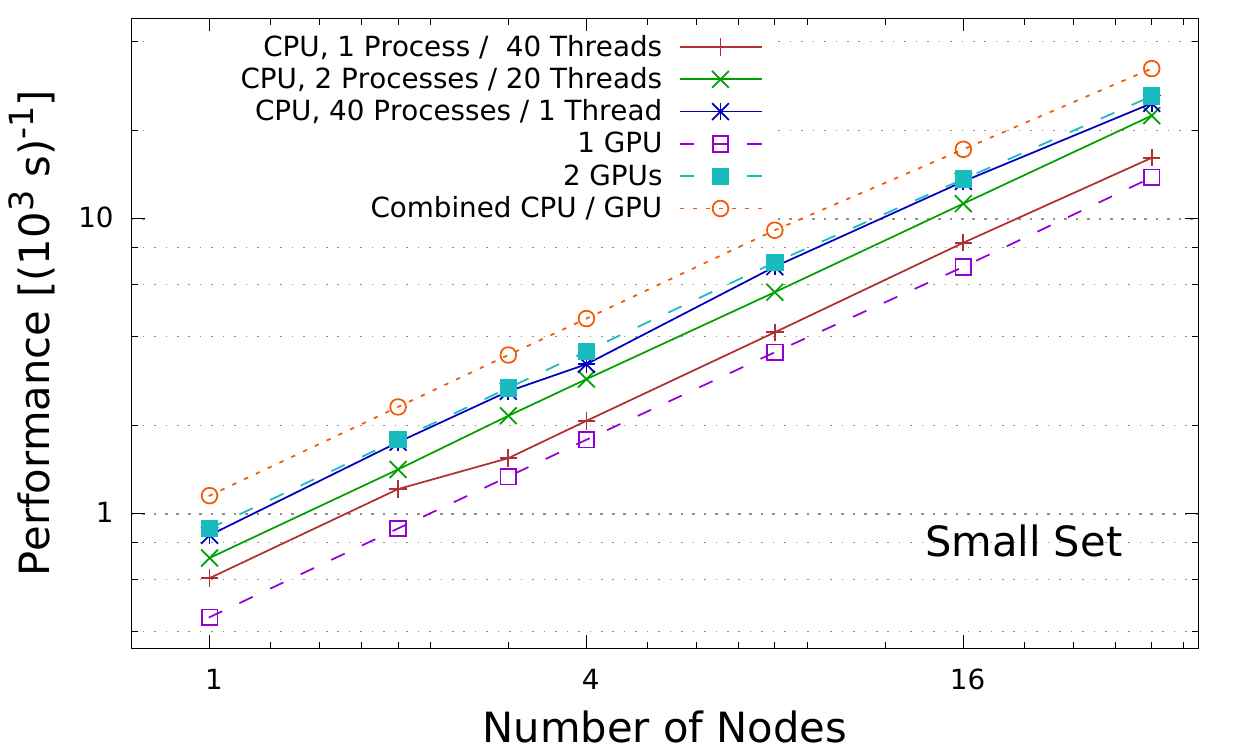}
\par\end{centering}
\caption{ BioEM wall-clock performance (inverse runtime) as a function of the number of compute nodes for the small dataset, with 100 observed images,
employing different parallelization and GPU-acceleration options.
Solid lines are CPU-only configurations, and squares with dashed lines are GPU-only configurations.
The combined workload configuration (open circles and dashed lines) had 60\% of the observed images on 2 GPU devices, and the remaining 40\% on the CPU with 20 OpenMP threads per node.
}
\label{fig:perf_small}
\end{figure}

\begin{figure}[!ht]
\begin{centering}
\includegraphics[width=\textwidth]{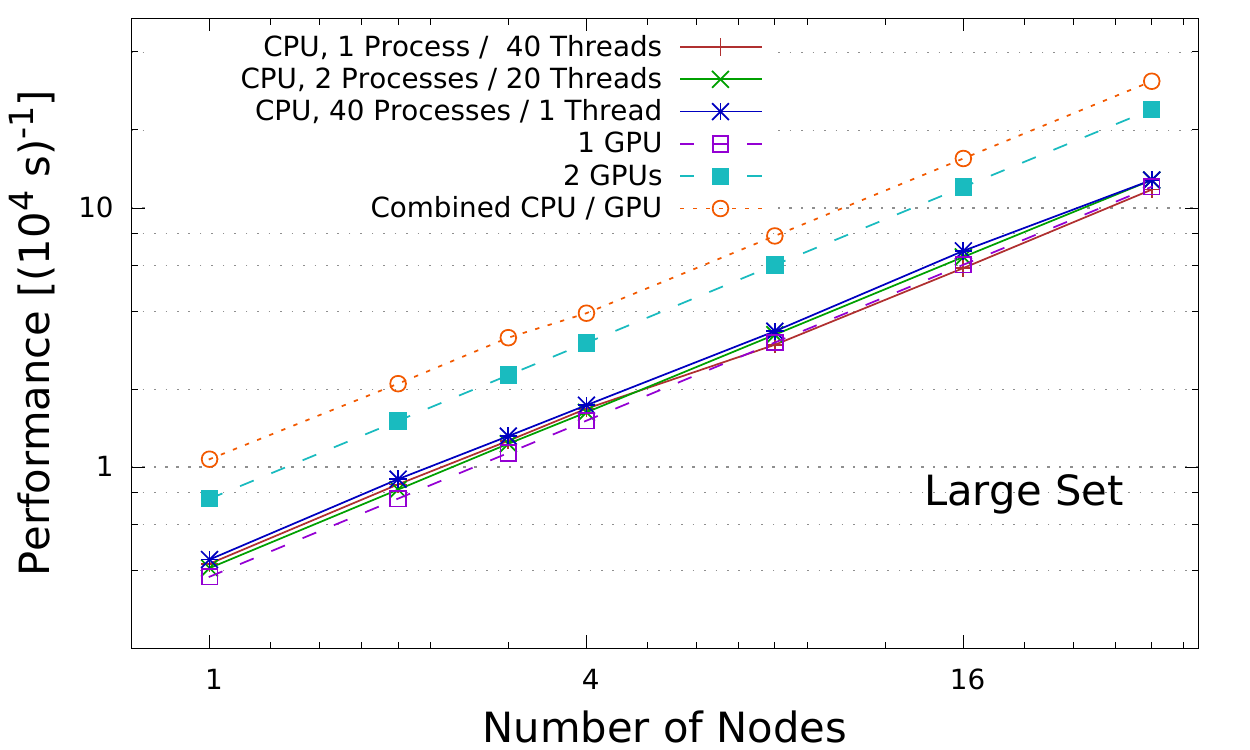}
\par\end{centering}
\caption{ BioEM wall-clock  performance (inverse runtime) as a function of the number of compute nodes for the large dataset, with 2000 observed images,
employing different parallelization and GPU-acceleration options.
Solid lines are CPU-only configurations, and squares with dashed lines are GPU-only configurations.
The combined workload configuration (open circles and dashed lines)  had 60\% of the observed images on 2 GPU devices, and the remaining 40\% on the CPU with 20 OpenMP threads per node.}
\label{fig:perf_large}
\end{figure}

Fig.~\ref{fig:perf_small} and~\ref{fig:perf_large} provide an overview of the performance, defined as the inverse runtime, 
obtained by employing different parallelization and GPU-acceleration options implemented by BioEM for the small 
and large datasets respectively on multiple nodes of the \emph{Hydra} cluster.
Both figures show multiple curves for different execution configurations.
The curves distinguish between CPU-only configurations, GPU-only configurations, and a combined configuration that uses both GPUs and all CPU cores of each node.
In the GPU-only case, two curves for one and for two GPUs per node are shown.
We present three curves in the CPU-only case that differ in the parallelization approach.
We show curves with pure OpenMP parallelization inside the nodes (1 process, 40 threads), pure 
MPI parallelization (40 single-threaded processes per node), and a hybrid MPI-OpenMP configuration of two MPI processes with 20 OpenMP threads each per node.
All configurations obtain a close-to-perfect linear scaling with the number of nodes due to the absence of both  communication and load-imbalances in the implementation.

The hybrid MPI-OpenMP parallelization adds flexibility for the parallelization and helps to contain the memory footprint.
It can reduce the number of MPI processes per node to only a few and multiple cores per MPI process can be used efficiently by the OpenMP parallelization.
For instance, assigning two MPI processes per two-socket node (i.\,e.~one per NUMA domain) avoids all NUMA limitations and reduces the memory footprint enormously 
(e.\,g., from 35~GB to around 2~GB per node for the large dataset).
On top of that, this shifts the strong scaling limit from the number of orientations to the number of orientations multiplied by the number of cores per socket.
Thus, the hybrid approach can speed up computations employing larger computational resources when the number of orientations limits the MPI 
scaling. 
In that respect, the new Intel Xeon Phi many-core CPU (codename:
``Knights Landing'') with about 240 threads and a scarce resource of 16 GB of high-bandwidth memory, 
presumably to be separated into four NUMA domains, might be a powerful and energy-efficient architecture 
for operating the BioEM software in the hybrid MPI-OpenMP setup.

The large dataset comprising more images allows the GPU a better exploitation of its parallel architecture.
For the small benchmark task, one GPU is as fast as one of the two 10-core Ivy-Bridge processors.
Processing the large dataset, the GPU runs twice as efficiently achieving the performance of a full Ivy-Bridge node with two processors.
(Consider that the small set with only 100 images represents more or less a lower bound, so GPU-processing works for all real cases.)
In both cases, two GPUs reach twice the performance of a single GPU.

The GPU achieves a smaller portion of its theoretical peak performance compared to the processor, because reading 
the images from global memory saturates the memory bandwidth.
Here, the processor can play the strength of its larger caches.
We note that whereas CPUs can easily hold tens of thousands of images, GPUs are slightly limited in memory ($\sim 6$\,GB for a K20x).
However, the limited memory size is no real restriction because processing the images takes place independently.
The GPU can process subsets of the images step-by-step, and the host can combine the results later on.
Since processing of such a subset of up to~$6$\,GB of images takes on the order of minutes, the overhead for additional transfers and repeated projections, etc., of the model is negligible.

BioEM also allows us to split the workload among CPU and GPU, i.\,e., the observed images are split into two sets: one goes to the GPU, and the other is analyzed on the CPU using OpenMP.
We find that this workload-sharing improves the performance significantly.
The full capacity of the node is utilized, profiting from the 2 GPU devices and all the cores in the node.
The optimal splitting ratio depends on the specific problem and hardware.
For these setups, the fastest setting was 60\% GPU for the small, and 65\% GPU for the large datasets.
Due to synchronization issues, the combined configuration does not achieve the sum of the individual CPU and GPU performances.
(For instance, the combined configuration on~32 nodes achieves 84\% of the sum of the individual performances of the large dataset.)

Our performance assessment shows that the optimal execution setup depends on the problem.
By trend, the MPI parallelization works better with many orientations, while OpenMP needs many observed images.
A hybrid setup is often the best compromise and scales almost perfectly linear with the number of cores.
For most cases, we recommend employing one MPI process per NUMA domain.

For instance, a complete analysis of 10000 images with 13500 orientations and 64 PSFs
takes approximately 140 minutes on 16 Ivy-Bridge nodes.
The same analysis can be performed within 55 minutes if the nodes are accelerated by two K20x GPUs.
This demonstrates that our new software can efficiently handle the analysis of the large amounts of experimental particles used in electron microscopy.
We estimate that per grid point (one PSF and one orientation), using the Fourier-algorithm,
the code takes approximately 0.1\,ms per image of $\sim$50000 pixels, with the exact runtime depending on the specific setup: number of pixels, 
model size, parameter ranges, grid points, etc.
In the Supplementary Information, we present some more estimates on the BioEM software runtime.

\section{Discussion}

The BioEM method provides an alternative approach to structurally characterize biomolecules using electron microscopy images.
By calculating the posterior probability of a model with respect to each individual image, it avoids information loss in averaging or classification,
and allows us to compare structural models according to their posterior probability.
Bayesian analysis methods, such as Relion~\cite{Scheres_JMB_12,Bayesian-Scheres}, have been enormously successful in EM, contributing much to the resolution
revolution~\cite{ResolRev_Kuelbrandt}. However, the main use of Relion is in reconstructing 3D densities 
from projection images, and not to rank or compare existing structural models. It also differs from BioEM in 
the integration scheme and optimization algorithms (for a comparison see Supplementary Table S1). 
BioEM requires relatively few images to discriminate the correct model within a pool of plausible structures 
(e.\,g., $<$1500 particles for GroEL~\cite{CossioHummerJSB_2013}), whereas, for full 3D reconstructions, 
Relion typically requires tens to hundreds of thousands of particles and costly computational resources to
 implement the multiple methods that select, classify, and polish the particles as well as refine the 3D maps.
Beyond applications in studies of highly dynamic systems~\cite{CossioHummerJSB_2013}, we envision that BioEM can complement 
traditional 3D reconstruction techniques in the first steps of classification by assigning accurate orientations 
and single-particle PSF estimations, and in the last steps of refinement by validating the final 3D models.
In addition, the BioEM method can be applied to problems where reconstruction techniques fail, e.\,g., when there are
few particle images that acquire preferred orientations or when the system is flexible.

The mathematical framework can also be extended to analyze individual time-dependent frames from direct electron-detection
cameras or electron tomography tilt-series  (see the Supplementary Information).
We foresee that the BioEM method can be generalized to other types of imaging experiments, such as atomic force or light microscopy 
after appropriate modifications of the forward calculation of $I^{\mathrm{cal}}$.

A possible limitation of the method is that structural models are required.
However, the models can be constructed using low-resolution data and hybrid modeling, e.\,g., by
combining coarse-grained maps with components from homologous PDB domains and models from simulations.
However, because the set of models is incomplete (the normalization in Eq.~\ref{eq:Pmom} is missing), 
BioEM cannot give an absolute estimation of the posterior probability but rather a relative value.
Thus, model comparison  is essential in the BioEM framework.

The BioEM software scales almost ideally with the number of CPU cores and has excellent performance on both CPU and GPU architectures.
The code has been optimized for a fast, and accurate analysis of tens of thousands of images, 
as is required in electron microscopy, and is sufficiently flexible to adjust to diverse research necessities.

In order to cope with the growing heterogeneity of GPU-accelerated
systems (in terms of number of CPUs and GPUs within a node and their relative
performance) we plan in the future to add an autotuning feature to BioEM which dynamically
chooses the optimal distribution of the workload between the CPUs and the GPUs
of a node. The distribution can be continuously adjusted based on measurement of the current CPU and GPU image processing rate.
Moreover, the autotuning could suggest a good setting for the number of MPI processes per node for the hybrid MPI-OpenMP mode.

The performance of BioEM is dominated to a great extent by the FFT libraries~\cite{FFTW05} ({\ttfamily fftw} on the CPU and {\ttfamily cuFFT} on the GPU, respectively), 
which are well optimized, leaving little margin for performance improvements.
Specifically, we run three consecutive steps for computing the cross-correlation 
and posterior probability: multiplication in Fourier space, fast Fourier backtransformation, and evaluation of the analytic formula.
The multiplication of the images in Fourier space can saturate the memory bandwidth of the GPU.
Since we use the  {\ttfamily cuFFT} library which performs the FFT as a black box, the intermediate data must be stored before and after the FFT.
However, we only need certain Fourier coefficients but  {\ttfamily cuFFT} computes all of them.
It could be possible to modify the  {\ttfamily cuFFT} to extract only the relevant coefficients. However, 
we have not pursued this yet considering the challenges required to modify and maintain such code.

A different optimization is more promising.
It is possible to parallelize over the projections via OpenMP instead of 
MPI using mutexes to control the probability updates of the observed images.
Such an implementation should achieve the same performance as the MPI version 
but keep the small memory footprint of the OpenMP version.
This would be optimal to analyze images where the molecular orientations are not distributed randomly 
but are correlated.
For example, in electron tomography correlations arise in different tilt
images of the same particle.

\section*{Acknowledgments}

The authors acknowledge Prof.~Dr.~K\"uhlbrandt, Dr.~Vonck and Dr.~Allegretti for the availability of the Frh 
micrographs, and feedback on the experimental techniques. P.C.,  F.B., M.R., and G.H. were supported by the Max Planck Society.
The BioEM software can be downloaded on the webpage:
\noindent https://gitlab.mpcdf.mpg.de/MPIBP-Hummer/BioEM,  jointly with a manual and tutorial.

\section*{References}
%\bibliographystyle{elsarticle-num}
%\bibliography{Biblio_BioEM}

\end{document}